\documentclass[aip,pof,floatfix,amsmath,amssymb]{revtex4-1}
\usepackage{amssymb}
\usepackage{bm}
\usepackage{url}
\usepackage{graphicx}

\usepackage[dvipsnames,usenames]{color}
\begin{document}

\title{Clustering and Turbophoresis in a Shear Flow without Walls}

\author{Filippo De Lillo}\email[corresponding author:]{delillo@to.infn.it} \affiliation{Dipartimento
  di Fisica and INFN, Universit\`a di Torino, via P. Giuria 1, 10125
  Torino, Italy}

\author{Massimo Cencini}
\affiliation{Istituto dei Sistemi Complessi, Consiglio Nazionale delle 
Ricerche, via dei Taurini 19, 00185 Rome, Italy}

\author{Stefano Musacchio}
\affiliation{Universit\'e Nice Sophia Antipolis, CNRS, Laboratoire J. A. Dieudonn\'e,  
UMR 7351, 06100 Nice, France}

\author{Guido Boffetta}
\affiliation{Dipartimento di Fisica and INFN, Universit\`a di Torino, 
via P. Giuria 1, 10125 Torino, Italy}

\begin{abstract}
We investigate the spatial distribution of inertial particles
suspended in the bulk of a turbulent inhomogeneous flow.  By means of
direct numerical simulations of particle trajectories transported by
the turbulent Kolmogorov flow, 
we study large and small scale mechanisms inducing inhomogeneities in the distribution of heavy particles. We discuss turbophoresis both
for large and weak inertia, providing heuristic arguments for the
functional form of the particle density profile. In particular, we
argue and numerically confirm that the turbophoretic effect is maximal
for particles of intermediate inertia. Our results indicate that small-scale fractal clustering and turbophoresis peak in different ranges in the particles' Stokes number and the separation of the two peaks increases with the flow's Reynolds number.
\end{abstract}


\maketitle 

\section{Introduction}
Turbulent aerosols, dilute solutions of solid particles
transported by turbulent flows, are important to the environment and
to industry. From combustion processes in coal fire burners, to the
dynamics of droplets in clouds, turbulent aerosols impact on our
life and the earth's climate \cite{hidy2012,Reeks2014}.  One general feature of
turbulent aerosols is their `unmixing' while transported by the flow,
which is relevant to several processes including: warm-rain initiation
\cite{Falkovich2002,Shaw2003}, planetesimal formation in the early
solar system \cite{weidenschilling1980,tanga1996,Bracco1999}, chemical reactions and
industrial processes \cite{williams1979,xiong1991}.  In recent years
much attention has been gathered by the dissipative dynamics resulting
from particle inertia which can induce small-scale fractal clustering
also in homogeneous flows
\cite{balkovsky2001,Falkovich2002,bec2003,Shaw2003,Duncan2005,bec2007}.
This can have relevant consequences for the rate of collision, coalescence 
and reaction of particles. Another well known unmixing mechanism in
turbulent aerosols is turbophoresis: inertial particles migrating in
regions of lower turbulent diffusivity, similarly to
thermophoresis~\cite{maxwell1878}, for which Brownian particles
are subject to an effective drift opposite temperature gradients. Turbophoresis has been mostly studied in presence of
boundaries, because as a mechanism for particle deposition in turbulent
boundary layers \cite{Sehmel1970,Brooke1992} it finds applications both
for industrial processes (for removing submicron sized particles from
gas streams) and the environment (dry deposition in the
atmosphere~\cite{Caporaloni1975}). Nonetheless, the mechanism of turbophoresis is independent of the presence of boundaries as, in principle, it only requires the presence of inhomogeneities in the flow. 

In this work we investigate the phenomenology of turbophoresis in a turbulent
shear flow without walls. 
We point out the differences between this mechanism which causes 
inhomogeneity at large scales and 
the small-scale clustering which occurs at viscous scales.

\section{Equations of Motion and Parameters}
As a paradigmatic case of inhomogeneous unbounded flow, we consider
the turbulent Kolmogorov flow, obtained by sustaining
the Navier-Stokes equations for the incompressible velocity field
$\bm u$,
\begin{equation}
\label{eq:NS}
\partial_t \bm u+{\bm u}\cdot\bm\nabla{\bm u}=-\bm\nabla p+\nu\Delta{\bm u}+{\bm F}(z)\,,
\end{equation}
with a sinusoidal force ${\bm F}(z)=F_0\cos(z/L)\hat{\bm e}_1$, where
$p$ is the pressure, $\nu$ the fluid kinematic viscosity, and
$\hat{\bm e}_1$ denotes the unit vector along the horizontal
direction.  
The laminar fixed point ($\bm u=U \cos(z/L)\hat{\bm e}_1$, with 
$U=L^2F_0/\nu$) becomes unstable above a critical Reynolds number \cite{Sivashinsky1985}, 
$Re=UL/\nu>\sqrt{2}$, and the flow eventually 
becomes turbulent for large $Re$ \cite{borue1996}.
A remarkable peculiarity of monochromatic forcing is that the resulting
mean velocity profile, 
$\langle {\bm u}\rangle=U\cos(z/L) \hat{\bm e}_1$, 
is monochromatic also in the turbulent regimes
\cite{borue1996, Musacchio2014}.  Above and in the following, the
brackets $\langle \cdots \rangle$ denote the average over $(x,y)$
and over time, while \mbox{$\overline f\equiv \int_0^{2\pi L}\langle f\rangle dz/(2\pi L)$}.
Due to the change of direction of the mean flow every half wavelength,
the Kolmogorov flow can be seen as an array of virtual channels
flowing in alternate directions without being confined by material
boundaries.

The dynamics of a small spherical particle is described by 
the Maxey-Riley equation~\cite{maxey1983}. 
Here, we focus on dilute suspensions of very small particles much heavier than
the fluid, whose dynamics is dominated by the Stokes drag.  
In this limit, the equations for the position ${\bm x}$ and velocity
${\bm v}$ of each particle simplify to 
\begin{eqnarray}
\label{eq:MR}
\dot{\bm x}&=&{\bm v}\\
\label{eq:MR2}
\dot{\bm v}&=&-\frac{1}{\tau} [{\bm v}-{\bm u}({\bm x},t)]
\end{eqnarray}
where $\tau=(2a^2\rho_p)/(9\nu\rho)$ is the Stokes time, $a$ and
$\rho_p$ are the particle radius and density, respectively while
$\rho$ denotes the fluid density.  Eqs.~(\ref{eq:MR}-\ref{eq:MR2}) assumes a Stokes
flow around the particle, implying that the particle's Reynolds
number must be very small: \mbox{$Re_p=|{\bm v}-{\bm u}|a/\nu\ll 1$}.

Particle inertia is commonly parametrized in terms of the Stokes
number $St=\tau/\tau_\eta $ based on the Kolmogorov time $\tau_\eta$,
i.e., the smallest characteristic time of a turbulent flow.  However,
turbophoretic effects are expected to be determined by large-scale
features of the flow, namely by the interplay between the advection
and the inhomogeneities of the eddy diffusivity~\cite{Belan2014}.  We
therefore introduce a particle inertia parameter $S=\tau / T$ by
normalizing the particle response time $\tau$ with the large-scale
eddy turnover time $T = E/\epsilon $, defined as
the ratio between the mean kinetic energy $E$ and the energy dissipation
rate $\epsilon$.
The parameter $S$ is the
analogous of $\tau_+=\tau {u^*}^2/\nu$, which is used in wall-bounded flows
to parametrize turbophoresis\cite{Sardina2012,Marchioli2002} in terms of the
friction velocity $u^*$.  This amounts to measuring times in wall
units, which control the scaling of inhomogeneities across the wall
region.

\begin{figure}[h!]
\centering
\includegraphics[height=0.4\textheight]{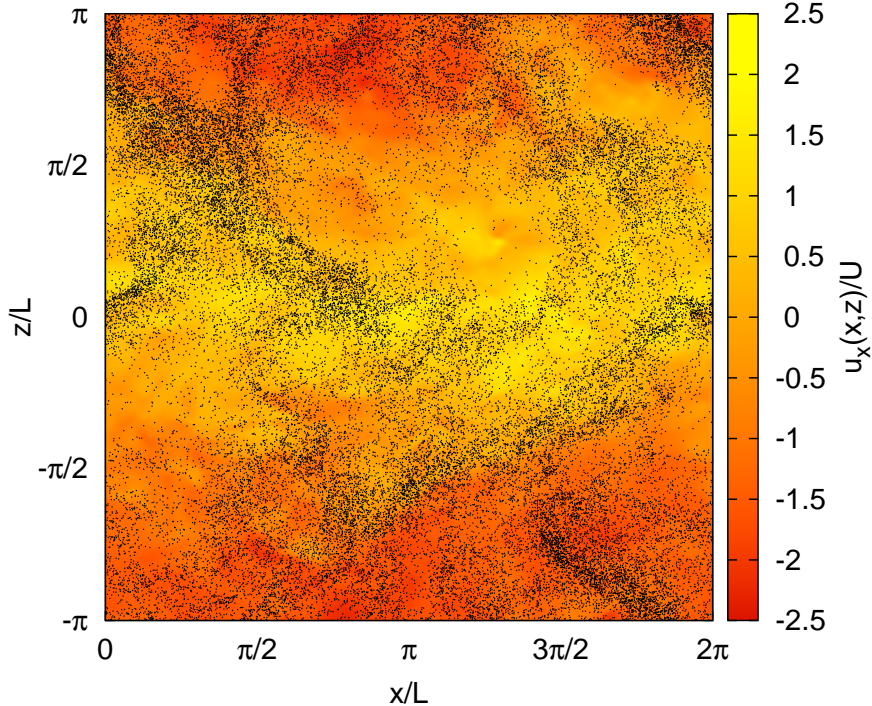}
\caption{Particle distribution in a slab of thickness $2\pi L/10$,
  plotted over the corresponding streamwise component of
  velocity (color map, red to yellow online). Particles inertia is $S=7.9\times10^{-2}$ and $Re=990$. 
}
\label{fig:1}
\end{figure}

\section{Results and discussion}
\subsection{Numerical Simulations}
We performed direct numerical simulations (DNS) of Eq.~(\ref{eq:NS})
by means of a standard pseudospectral code with triple-periodic
boundary conditions in a cubic domain of side $L_x=L_y=L_z=2 \pi$ at
resolution $N^3$, with $N=128$ and $256$ .  For each class of particles with given inertia $S$,
we integrated $4\cdot 10^5$ trajectories according to Eqs.~(\ref{eq:MR}-\ref{eq:MR2}),
with the fluid velocity obtained by linear interpolation from grid nodes
to particle positions.  Eulerian and Lagrangian dynamics is integrated
via a second-order Runge-Kutta scheme.  DNS parameters are reported
in Table~\ref{table1}.
\begin{table}[t!]
\begin{tabular}{cccccccc}
\hline \hline
$N$ & $\nu$ & $L$ & $F$ & $U$ & $T$& $\epsilon$ & $Re$ \\ \hline
$128$ & $1 \times 10^{-3}$ & $1.0$ & $8 \times 10^{-3}$ & $0.23$ &
$36.3$ &
$9.31 \times 10^{-4}$ & $230$ \\
$256$ & $1 \times 10^{-3}$ & $1.0$ & $1.28 \times 10^{-1}$ & $0.99$ &
$9.36$ &
$6.41 \times 10^{-2}$ & $990$\\
\hline \hline
\end{tabular}
\caption{DNS parameters: $N$ resolution, $\nu$ kinematic viscosity,
  $L$ forcing scale, $F$ forcing amplitude, $U$ amplitude of mean
  velocity profile, $T$ large-scale time, $\epsilon$ energy dissipation
  rate, Reynolds number $Re=UL/\nu$.}
\label{table1}
\end{table}

\begin{figure}[h!]
\centering
\includegraphics[height=0.23\textheight]{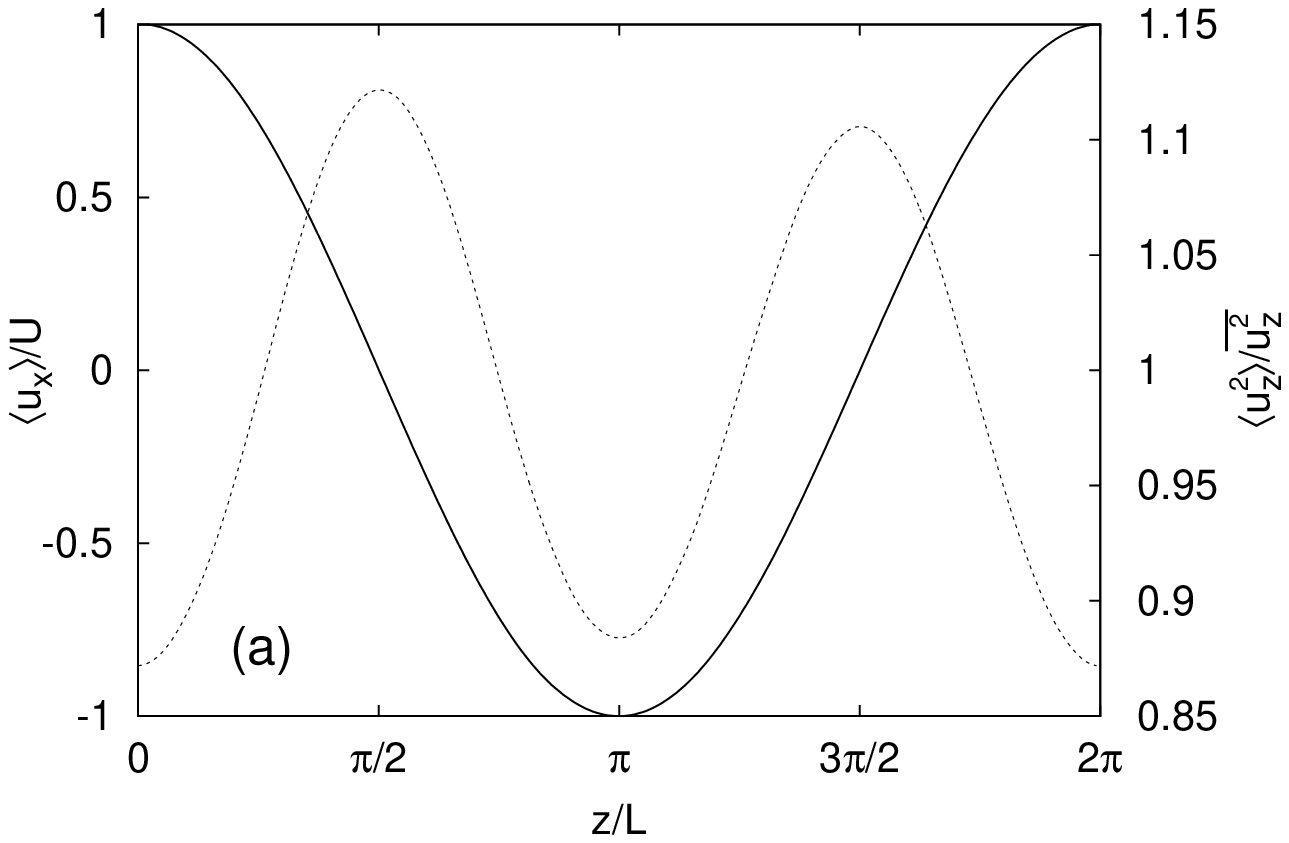}
\includegraphics[height=0.23\textheight]{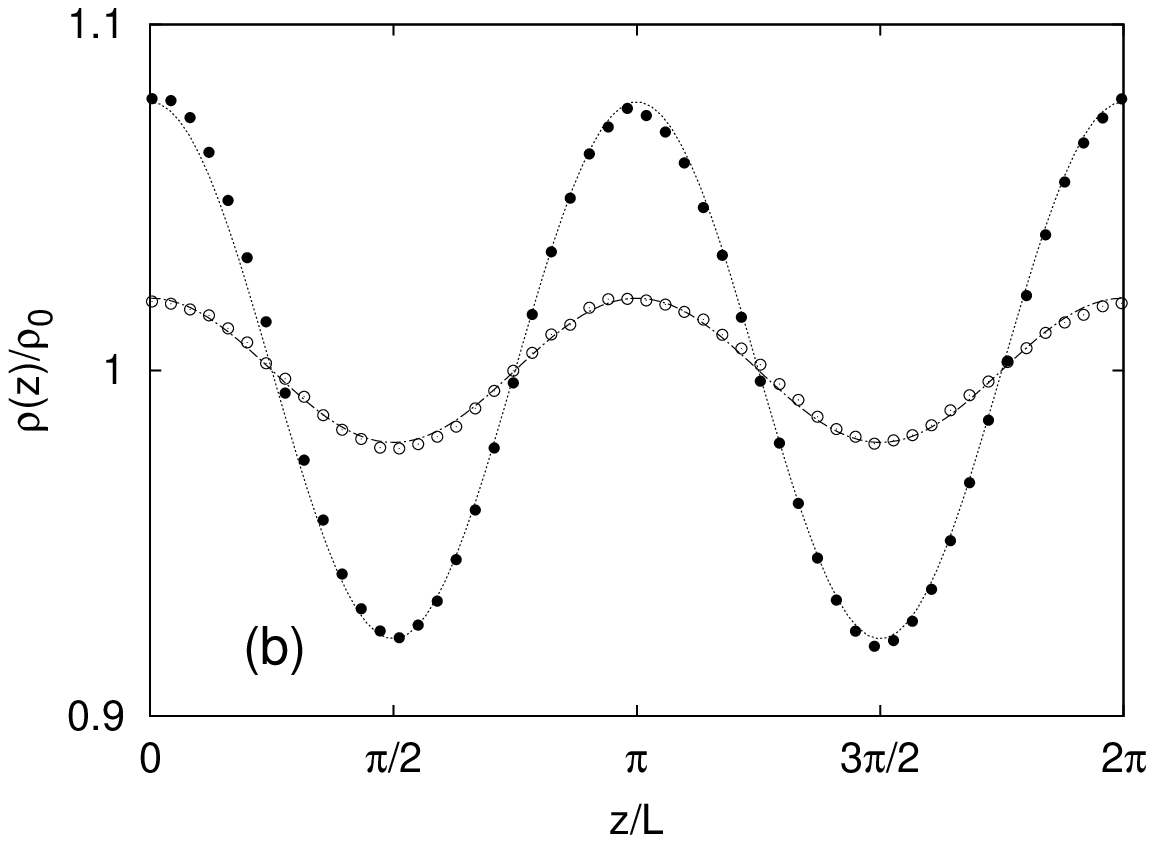}
\caption{
Fluid velocity and particle distribution profiles at ${\rm Re}=990$.
(a) profiles of the longitudinal velocity 
$\langle u_x \rangle$ (solid line, left axis) 
and the fluctuations in shear-normal kinetic energy 
$\langle u_z^2 \rangle$ (dashed line, right axis) of the flow. The small asymmetry in $\langle u_z^2\rangle$ is due to the finite statistics.
(b) particle number density profiles $\rho(z)$ for 
${\rm S}=7.9\times10^{-2}$ (filled circles) 
and 
${\rm S}=2.6\times10^{-3}$ (empty circles),
compared with the functional form~(\ref{eq:profile}) (lines, fitted).}  
\label{fig:2}
\end{figure}

Large-scale inhomogeneities are clearly visible in the particle
distribution in Fig.~\ref{fig:1}.  To reveal the correlations of particle
positions with the shear-normal structure of the flow it is necessary to
consider statistically averaged quantities. Figure~\ref{fig:2} shows typical
fluid velocity and particle number-density profiles obtained by averaging over
the $x$ and $y$ directions (normal to the shear) and over very long integration
of hundreds of
large-eddy-turn-over-times.  The modulation of the density profiles 
closely reflects the structure of the mean flow: particles concentrate 
in the regions of maximal mean flow and minimal mean shear, away from 
the maxima of turbulent energy.

As it is shown in (Fig.~\ref{fig:2}), the particle density profiles 
are accurately fitted by:
\begin{equation} 
\label{eq:profile}
\rho(z)=\rho_0(1+a(S)\cos(2z/L))
\end{equation}
where $\rho_0=1/L_z$ is the mean uniform density and the only
free-parameter is $a(S)$, which accounts for the dependence on the
particles' inertia.  In the following we discuss a heuristic argument
which gives support to the empirical formula (\ref{eq:profile}).

\subsection{Turbophoresis}
A common approach to derive theoretical predictions for the dynamics
of inertial particles is by modeling the velocity field as a
Gaussian, short-correlated noise \cite{balkovsky2001}. With this
assumption, one can write\cite{Belan2014} a Fokker-Planck equation for the
probability density $P(z,v)$ to find a particle in $z$ with vertical
velocity $v$, in which turbulence 
is parametrized by a space-dependent eddy diffusivity
$\kappa(z)$ acting on velocity and derived from Eq.~(\ref{eq:MR2}).
It is then possible, in the limit of fast relaxation of the velocity
distribution~\cite{Caporaloni1975,Belan2014}, to obtain
an equation for the marginal distribution $\rho(z)=\int dv P(z,v)$,
which reads $\partial_t \rho(z)=\partial_z J(z)$, where the flux is
$J(z)= \partial_z [\kappa(z) \rho]$. For the fluxless steady state
one obtains the prediction $\rho\sim \kappa^{-1}(z)$ which, in analogy
with thermophoresis~\cite{Lopez2007}, implies that particles
concentrate in the minima of diffusivity. This behaviour is
substantially different from that of a classical, passively advected scalar
field $\theta$, where the eddy-diffusivity would appear in the flux in the Fickian form 
$J=\kappa(z)\partial_z\theta$, leading to a homogeneous steady state. Standard
dimensional
arguments suggest that the eddy diffusivity is proportional to the
mean square velocity $\kappa(z) \propto \tau_c \langle u_z^2 \rangle$
(with $\tau_c$ an appropriate correlation time), so that the above
result implies $\rho(z) \propto \langle u_z^2 \rangle^{-1}$.  In the
case of the Kolmogorov flow, the profile of the mean square vertical
velocity is found to be $\langle u_z^2 \rangle \propto
U^2(1-b\cos(2z/L))$, with $b\ll1$ and weakly depending on Re~\cite{Musacchio2014}.  Using a first-order Taylor expansion in $b$ one
recovers the expression (\ref{eq:profile}).  It is worth remarking
that the above argument relies on two assumptions.  First, the
correlation time of the flow is set to zero.  Second, the particle
Stokes time $\tau$ is assumed to be small enough to justify the fast
relaxation of the velocity distribution.  In this limit the amplitude
of the spatial modulation of the particle density profile would not depend on $S$,
namely $a(S)=b$. The latter, quantitative prediction does not hold if the flow
has a finite correlation time, as in our case. However, we find that
Eq.~(\ref{eq:profile}) gives the correct shape for the density profile for
particles with Stokes
times both shorter and longer than the correlation time of the flow,
provided that the amplitude $a(S)$ is allowed to depend on inertia.

\begin{figure}[h!]
\centering
\includegraphics[height=0.3\textheight]{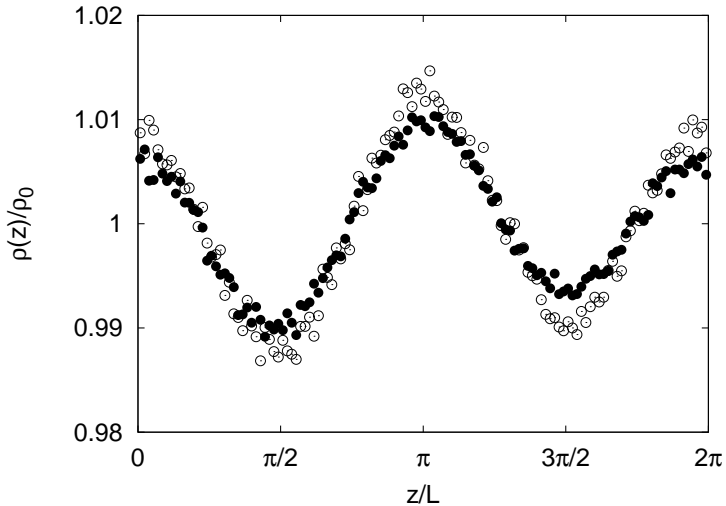}
\caption{Particle number density profiles $\rho(z)$ for 
${\rm S}=4.1$ (filled circles) at ${\rm Re}= 230$. 
The particle distribution is compared the prediction 
$\rho(z)/\rho_0 =  \langle v_z^2 \rangle^{-1} L_z/\int_0^{L_z} \langle v_z^2\rangle^{-1}dz$ 
(empty circles). Statistical fluctuations are due to the slow
convergence observed for large $S$. 
}
\label{fig:3}
\end{figure}

The analogy with thermophoresis can be exploited for particles with
large inertia. In this limit, the particles can be seen as a gas in
equilibrium with the turbulent environment and we can interpret the
spatial variations of the mean particle vertical velocity variance,
$\langle v_z^2(z) \rangle$, as the analogous of a space-dependent
temperature field \cite{Caporaloni1975}.  Assuming the local
diffusivity proportional to the temperature, i.e. $\kappa(z) \sim 
\langle v_z^2(z) \rangle$, the particle density profile is therefore
expected to be $\rho(z) \propto \langle v_z^2 \rangle^{-1}$, which is
in fairly good agreement with numerical results for large $S$ (see
Fig.~\ref{fig:3}).  Moreover, we find that the particle velocity
profile $\langle v_z^2 \rangle$ has the same spatial dependence as the
fluid one $\langle u_z^2 \rangle$, but the amplitude of the spatial
modulation decreases at increasing inertia.  This leads to the
prediction that the amplitude $a(S)$ in~(\ref{eq:profile}) is a
decreasing function of the inertia for large $S$.

The scenario is different for particles whose Stokes time is of the
order of the eddy-turn-over times in the inertial range of turbulence.
Such particles are able to follow only turbulent eddies of size $\ell$
with a turn-over time, $\tau_\ell$, longer than their Stokes time,
i.e. $\tau_\ell>\tau$.  Smaller eddies still act as a colored noise
giving raise to a space-dependent effective diffusivity responsible for
turbophoresis.  Conversely, eddies with $\tau_\ell>\tau$ mix the
particles almost like tracers, thus reducing the turbophoretic
accumulation.  Turbophoretic unmixing is therefore enhanced as $S$
increases, because a larger fraction of eddies contribute to it.
Assuming that the profile of the effective diffusivity due to the
small eddies has a monochromatic modulation one recovers the
prediction~(\ref{eq:profile}) in which $a(S)$ increases with $S$ for
small values of $S$.  Hence, we expect that $a(S)$ attains its maximum
when the particle response time is of the same order of the
characteristic time of the large-scale structures of the flow
($S\simeq O(1)$).

At the heart of the arguments discussed above, 
there is the notion that turbophoresis 
drives particles away from the maxima of turbulent energy,
which correspond to maxima of the eddy diffusivity.
In the case of the Kolmogorov flow, 
the maxima of turbulent fluctuations occur
where the shear of the mean flow is maximum and the mean flow
vanishes, i.e., at the borders between the virtual channels. 
Therefore, particles are driven toward the center of the virtual channels.  
This is in contrast with the case of a  turbulent channel (or pipe)
flow, in which turbulence is intense in the bulk and
vanishes in the viscous sub-layer close to the walls. In this case
turbophoresis drives the particles away from the bulk and 
concentrates them along the walls \cite{mclaughlin1989,Kaftori1995,Sardina2011}.
In this sense, the fact that turbophoresis may eventually
accumulate the particles to regions of large or small mean velocity
(or mean shear) is an incidental (albeit relevant for applications) 
consequence of the details of the particular flow considered. 

\begin{figure}[h!]
\centering
\includegraphics[height=0.3\textheight]{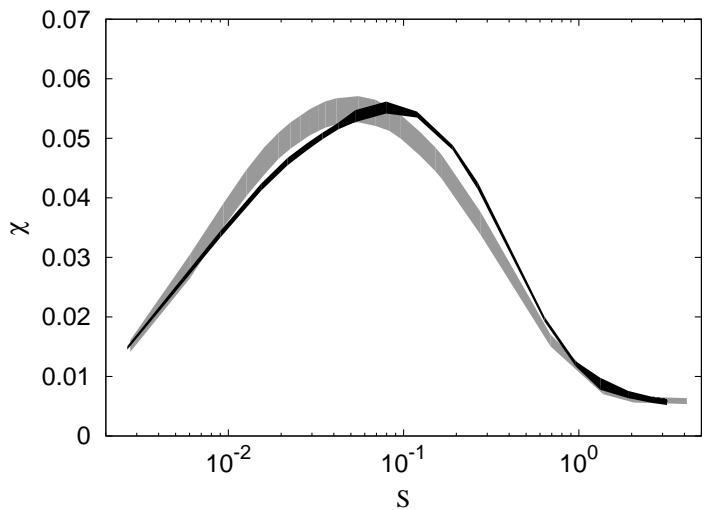}
\caption{Rms relative deviation $\chi$ 
from the homogeneous distribution 
plotted as a function of $S$ for $\rm{Re}=230$ (grey curve) 
and $\rm{Re}=990$ (black curve). The thickness of the curves reflects 
the statistical fluctuations around the mean value.
} 
\label{fig:4}
\end{figure}

The overall effect of turbophoresis can be quantified by means of the
rms relative deviation of the mean density profile $\rho(z)$ from the
uniform distribution $\rho_0$ as $\chi=[ 1/L_z
  \int_0^{L_z}(1-\rho(z)/\rho_0)^2 dz ]^{1/2}$.  For the specific
profile~(\ref{eq:profile}), clearly we have $\chi(S) = a(S)/\sqrt{2}$.
This quantity is plotted in Fig.\ref{fig:4} as a function of the
inertia parameter.  
In agreement with expectations, the turbophoretic
effect is not monotonic as a function of inertia. It displays a
maximum at $S \simeq 10^{-1}$.  The shape of the curves is not
strongly affected by changing $Re$ even though we observe, within the
statistical uncertainties, a weak dependence of the position of the
maximum.

Remarkably, deviations from the uniform distribution are present also
for particles whose Stokes time is much smaller than the Kolmogorov
time.  Arguments based on local variations of the eddy diffusivity can
not be used to explain the origin of such inhomogeneities, because the
particle relaxation time is shorter that the shortest eddy-turnover
time of the flow. The mechanism responsible for such
inhomogeneities also for $St \ll 1$ is related to the weak
compressibility of the particle velocity field.  When $St \ll 1$,
expanding at first order in $\tau$ the velocity of the particle one
has ${\bm v} = {\bm u} - \tau \left(\partial_t {\bm u}+{\bm u}
\cdot \nabla {\bm u}\right) + o(\tau)$
(see e.g. Ref.~\cite{balkovsky2001}).  The mean vertical profile of
the divergence of the particle velocity field is $\langle \nabla \cdot
{\bm v} \rangle = - \tau \langle \cdot \nabla ({\bm u} \cdot \nabla
{\bm u})\rangle = - \tau \partial_z^2 \langle u_z^2 \rangle$.  The
mean divergence is positive in the maxima of $\langle u_z^2 \rangle$ and is
negative in the minima, providing an explanation for the
accumulation of inertial particles in the minima of $\langle u_z^2
\rangle$, observed at very weak inertia.

\subsection{Small-Scale Clustering}
\begin{figure}[t!]
\centering
\includegraphics[height=0.3\textheight]{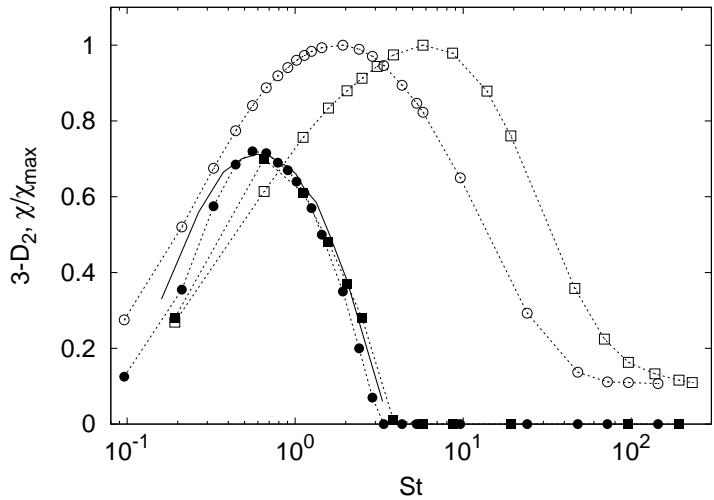}
\caption{Correlation co-dimension $3-D_2$ of particle distributions
(filled symbols) and rms relative deviation from uniform
distribution $\chi$ (empty symbols, same data of Fig.~\ref{fig:4}) 
as a function of $St$ (filled
symbols) and  for $Re=230$ (circles) and $Re=990$ (squares). The continuous
line show the comparison with the $D_2$ computed for a HIT case from
\cite{bec2007} at $Re_\lambda=185$.
} 
\label{fig:5}
\end{figure}

Besides the large-scale effects discussed above, 
inertial particles transported in a turbulent flow display small-scale clustering.  Small-scale spatial inhomogeneities
originate from the dissipative dynamics in the $6$-dimensional
position-velocity phase space $({\bm x},{\bm v})$
\cite{balkovsky2001,bec2003}.  In particular, inertial particle
motion asymptotically takes place on a (multi-)fractal set in phase
space. A fractal dimension smaller than space dimension
signals an enhanced probability to find particle
pairs at short separation. Indeed, the probability to find particle pairs at separation below a certain $r$ (smaller than the Kolmogorov scale)  
 grows as $r^{D_2}$, with $D_2=3$ for uniformly distributed particles
in three dimensions \cite{pv87}. The correlation dimension $D_2$ is
thus commonly used as a measure of clustering.
In Fig.~\ref{fig:5} we plot the co-dimension $3-D_2$ as a function of
$St=\tau/\tau_\eta$ for two values of $Re$.  In agreement with
previous results obtained in homogeneous, isotropic turbulence
(HIT)\cite{bec2007}, we find that the fractal co-dimension has only a
very weak dependence on Re. Moreover, we find that it is not affected
by the large-scales inhomogeneities of the Kolmogorov flow
\cite{Note1}
as apparent from Fig.~\ref{fig:5} where published data for $D_2$ 
of heavy particles from a HIT simulation \cite{bec2007} are shown for
comparison\cite{Note2}.
On the contrary, the turbophoretic clustering measured by $\chi$ plotted as a
function of $St$ has a strong dependence on $Re$: the maximum is
attained for larger $St$ as $Re$ increases (see Fig.\ref{fig:5}).

The different $Re$-dependence of the two phenomena reflects 
their different nature.  The small-scale clustering is due to the
chaotic dynamics at viscous scales, therefore it is most effective at
$St\sim O(1)$, i.e. when $\tau\simeq\tau_\eta$.  Conversely,
turbophoresis is the result of the transport of particles across the
large-scale inhomogeneities of the flow.  Its effects is maximum for
particles with response time of the order of the large-scale eddy
turnover time $T$, i.e, at $S \simeq 1$.  As the Reynolds number
grows, the scale separation between the two unmixing mechanisms is
expected to grow as $T/\tau_\eta\sim Re^{1/2}$, as shown in Fig.~\ref{fig:5}.

\section{Conclusions}
We have investigated the phenomenon of turbophoresis and fractal
clustering of heavy inertial particles in the bulk of inhomogeneous
flow, by performing DNS of the dynamics of
heavy particles transported by the turbulent Kolmogorov flow.  The
emerging scenario in the limit of large Re is the following.  The
distribution of particles with small inertia is characterized by a
strong fractal clustering at small scales, but is weakly affected by
turbophoresis. On the contrary, particles with large inertia
experience strong turbophoretic accumulation at large scales while
remaining uniformly distributed at small scales.  
The turbophoretic effect is maximum for
particles with Stokes time of the order of the large-scale
eddy-turnover times of the turbulent flow. Conversely, small-scale
fractal clustering is maximal for particles with Stokes times comparable with the Kolmogorov time. 

Turbophoresis is
characterized by large scale particle density profiles which are
strongly correlated to the inhomogeneities of the flow.  In particular,
particle density is maximal in the minima of the turbulent eddy
diffusivity, which for the case of the Kolmogorov flow coincides with
the maxima of the mean flow.  
This is an important difference with what observed in wall bounded flows, 
where turbophoresis concentrates particles in regions of minimum mean flow
close to the boundaries and demonstrates that the regions of particle
accumulation depend on the details of the flow.

\begin{acknowledgments}
We acknowledge support from the European COST Action MP1305 ``Flowing Matter''. Numerical simulations were performed at CINECA via the INFN-FieldTurb grant.
\end{acknowledgments}


%

\end{document}